\documentclass[conference]{IEEEtran}
\pdfoutput=1
\newtheorem{definition}{Definition}[section]

\usepackage{enumitem}

\usepackage{times}

\usepackage[T1]{fontenc}
\usepackage[scaled=0.81]{beramono}

\newcommand{\e}[1]{\mbox{\lstinline|#1|}}

\usepackage[pdftex]{graphicx}
\usepackage{amsmath,amssymb}
\usepackage{xcolor}
\usepackage{booktabs}
\usepackage{colortbl}
\usepackage{xspace}

\usepackage{listings}
\usepackage{eiffel}
\usepackage{boogie}

\usepackage{zed-csp}

\ifCLASSINFOpdf
\else
\fi

\begin{document}
%
\title{Complete contracts through specification drivers}

\author{\IEEEauthorblockN{Alexandr Naumchev\IEEEauthorrefmark{1},
Bertrand Meyer\IEEEauthorrefmark{2}}
\IEEEauthorblockA{Software Engineering Laboratory\\
Innopolis University\\
Innopolis, Russian Federation\\
\IEEEauthorrefmark{2}Also Politecnico di Milano\\
Email: \IEEEauthorrefmark{1}a.naumchev@innopolis.ru,
\IEEEauthorrefmark{2}Bertrand.Meyer@inf.ethz.ch
}
}


%


\maketitle

\begin{abstract}
Existing techniques of Design by Contract do not allow software developers to specify complete contracts in many cases. Incomplete contracts leave room for malicious implementations. This article complements Design by Contract with a simple yet powerful technique that removes the problem without adding syntactical mechanisms. The proposed technique makes it possible not only to derive complete contracts, but also to rigorously check and improve completeness of existing contracts without instrumenting them.
\end{abstract}

\begin{IEEEkeywords}
specification driver, abstract data type, Design by Contract, complete contract, Eiffel, Hoare triple, AutoProof
\end{IEEEkeywords}


%
\IEEEpeerreviewmaketitle

\section{Introduction}
\label{sec:introduction}
The main contribution of this work is a new approach to seamless software development, bridging the heretofore wide gap between two fundamental and widely used techniques: Abstract Data Types (ADTs) and Object-Oriented Programming (OOP). These techniques seem made for each other, but trying to combine them in practice reveals a glaring impedance mismatch. We explain the problem, provide a remedy, and subject it to formal verification.

ADTs~\cite{guttag1977abstract} are a clear, widely known way to specify systems precisely. OOP~\cite{meyer1988object} is the realization of ADT ideas at the design and programming level, with Design by Contract (semantic properties embedded in the program) providing the connection. At least, that is the accepted view. However, the correspondence is far less simple than this view would suggest. While it would seem natural to use ADTs for specification and OOP for design and implementation, in practice this combination hits an impedance mismatch:

\begin{itemize}
\item At the ADT level, some axioms involve two or more commands. For example, an axiom for stacks (the standard example of ADTs, which remains the best for explanatory purposes) will state that if you push an element onto a stack and then pop the stack, you end up with the original stack.

\item In a class, the standard unit of OOP, the contracts can only talk about one command, such as push or pop, but not both. Specifically, the postcondition of a command such as push can describe the command's effect on queries such as top (after you have pushed an element, that element is the new top), but there is no way to refer to the effect on pop as expressed by the ADT axiom.
\end{itemize}

The present work introduces a practical solution to this mismatch. The essence of the solution is that classes directly reflecting ADTs, such as a class STACK, cannot by themselves capture such multi-command (or "second-degree") ADT axioms, but this does not mean that the OOP approach fails us. The idea will be to introduce auxiliary classes whose role is to "talk about" the features of the basic classes such as STACK (the ones directly corresponding to ADTs). Such a class has features that combine those of basic classes, e.g. a command \e{push_then_pop} that works on an arbitrary stack, pushing an element on a stack and then popping the stack. Then the postcondition of \e{push_then_pop} can specify that the resulting stack is the same as the original.

We call such features specification drivers by analogy with "test drivers", which are similarly added to the basic units of a system for the sole purpose of testing them. Like test drivers, specification drivers serve purely verification purposes, rather than providing system functionality. The difference is of course that test drivers appear in dynamic verification (testing), whereas specification drivers are for static verification (for example, as in this paper, correctness proofs). But the basic idea is the same.

Specification drivers are not just a specification technique; we also submit them to formal, mechanical verification. As part of the AutoProof formal verification tool~\cite{tschannen2015autoproof}, we have mechanically proved the correctness of the examples given in this paper.

 \autoref{sec:motivating_example} explains the problem through a working example. \autoref{sec:details} describes the essentials of the solution. \autoref{sec:related_work} compares this approach with other possible ones. \autoref{sec:proving_completeness} presents our experience with mechanical verification. \autoref{sec:conclusions} draws conclusions and outlines future research prospects.

\section{Motivating Example}
\label{sec:motivating_example}

\begin{figure}[!b]
\textbf{TYPES}
\begin{itemize}
\item $STACK[G]$
\end{itemize}
\textbf{FUNCTIONS}
\begin{itemize}
\item $extend: STACK[G] \cross G \fun STACK[G]$
\item $remove: STACK[G] \pfun STACK[G]$
\item $item: STACK[G] \pfun G$
\item $is\_empty: STACK[G] \fun BOOLEAN$
\item $new: STACK[G]$
\end{itemize}
\textbf{AXIOMS} \\
\t1 For any $x: G$, $s: STACK[G]$
\begin{enumerate}[label=(A\arabic*)]
\item $item(extend(s,x))=x$
\item $remove(extend(s,x))=s$
\item $is\_empty(new)$
\item \textbf{not} $is\_empty(extend(s,x))$
\end{enumerate}
\textbf{PRECONDITIONS}
\begin{enumerate}[label=(P\arabic*)]
\item $remove(s:STACK[G])$ \textbf{require not} $is\_empty(s)$
\item $item(s:STACK[G])$ \textbf{require not} $is\_empty(s)$
\end{enumerate}
\caption{ADT specification of stacks}
\label{fig:stacks_adt}
\end{figure}

\autoref{fig:stacks_adt} contains the standard ADT specification of stacks. The standard names of the functions are changed in favor of the mechanical verification experiment in \autoref{sec:proving_completeness}: the existing implementation, to which the experiment is applied, uses exactly these names.

\begin{figure}[!b]
\begin{lstlisting}
class STACK_IMPLEMENTATION [G] -- Type STACK[G]
create	new -- Marking new as a creation feature
feature
	extend (x: G)  -- Extending with a new element
		do
		ensure
			a1: item = x
			a4: not is_empty
		end

	remove -- Removing the topmost element
		require
			p1: not is_empty
		do
		end

	item: G    -- The topmost element
		require
			p2: not is_empty
		do
		end

	is_empty: BOOLEAN  -- Is the stack empty?

	new    -- Instantiating a stack
		do
		ensure
			a3: is_empty
		end
end
\end{lstlisting}
\caption{Applying the traditional process of DbC to the stacks ADT specification}
\label{fig:stacks_traditional_approach}
\end{figure}

\autoref{fig:stacks_traditional_approach} contains the result of applying the traditional process of DbC~\cite{meyer1988object} to the specification in \autoref{fig:stacks_adt}:
\begin{itemize}
\item The name of the class is derived from the name of the ADT it implements.
\item The signatures of the implementation features are derivatives of the ADT functions' descriptions.
\item Preconditions of the ADT functions go to \textbf{require} clauses of the implementation features.
\item Postconditions of the implementation features capture ADT axioms A1, A3 and A4.
\item The \textbf{create} clause lists the implementation feature \e{new} to highlight its special mission of instantiating new stacks.
\end{itemize}
Axiom A2 introduces the problem. The axiom constrains two functions simultaneously, $extend$ and $remove$: the former one should do nothing but extend the stack with the given element, and the latter should do nothing but remove the topmost element of the stack. As a consequence, it is not possible to capture the axiom in a single implementation feature postcondition. Postconditions operate on two objects: the target object before calling the feature and the target object after invoking the feature. If the feature has formal parameters, they also parameterize the postcondition. Axiom A2 involves three stacks: the original one $s$, $s_1$ resulting from applying function $extend$ to $s$, and finally $s_2$ resulting from applying $remove$ to $s_1$. Formally:
\begin{align*}
\forall s, s_1, s_2: &\ STACK[G]; x: G @ \\
                     & (s_1 = extend (s, x) \land s_2 = remove (s_1) \implies s_2 = s
\end{align*}
Or, writing the quantified expression in terms of postconditions:
\begin{equation}
\label{eq:a2_post}
(Post_{extend}(s, s_1, x) \land Post_{remove} (s_1, s_2)) \implies s_2 = s
\end{equation}

On one hand, it is not possible to capture A2 in a single postcondition. On the other hand, postconditions of \e{extend} and \e{remove} should exist and be strong enough to satisfy \autoref{eq:a2_post}.

Failures to capture such important properties as A2 in postconditions leave room for invalid implementations. In particular, inability to capture axiom A2 makes it possible to implement stacks which store only the last added element and thus are useless as data containers. Still, such an implementation satisfies all the other axioms as its postconditions capture them.
\begin{figure}[!t]
\begin{lstlisting}
class STACK_IMPLEMENTATION [G]
create new
feature
	extend (x: G)	-- Extending with a new element
		do
			item := x
			is_empty := False
		ensure
			a1: item = x
			a4: not is_empty
		end

	remove	-- Removing the topmost element
		do
			 is_empty := True
		end

	item: G	-- The topmost element

	is_empty: BOOLEAN -- Is the stack empty?

	new	-- Instantiating a stack
		do
			is_empty := True
		ensure
			a3: is_empty
		end

end
\end{lstlisting}
\caption{Underspecified postconditions may lead to invalid implementations}
\label{fig:malicious_stack}
\end{figure}

\autoref{fig:malicious_stack} depicts such an invalid implementation. For the sake of simplicity, it ignores preconditions, but this does not render the reasoning invalid: an empty precondition defaults to \e{TRUE}, the weakest conceivable precondition. According to the rule of consequence for preconditions~\cite{hoare1969axiomatic}, correctness against a weaker precondition implies correctness against a stronger one. Submitting the class \e{STACK_IMPLEMENTATION} to AutoProof confirms the point: the tool successfully proves "correctness" of the implementation.

For purist developers the problem of underspecified postconditions may easily become a reason for not using them at all. Intuitively, it seems better to keep all the properties written in a single place, and the described problem prevents doing this: although it is possible to capture some ADT axioms in postconditions, some of them will have to exist in separate documents and thus carry the risk of misuse and all the associated traceability costs.

\section{Axioms as Specification Drivers}
\label{sec:axioms_as_specification_drivers}
The example in \autoref{fig:stacks_traditional_approach} translates axiom A1 directly to the postcondition of the implementation feature \e{extend}. Is it in fact the only way to do the translation of the axiom? A closer look at the original axiom and its translation in \autoref{fig:stacks_traditional_approach} reveals two facts:
\begin{itemize}
\item The axiom uses the function $extend$ in a sense of applying it, while its translation in \autoref{fig:stacks_traditional_approach} specifies the implementation feature directly without invoking it.
\item The axiom uses an explicit stack instance $s$, while the translation implicitly operates on the current object described by class \e{STACK_IMPLEMENTATION[G]}.
\end{itemize}
Is it possible to devise a translation of axiom A1 that would be closer to the origin?

Existing techniques of DbC completely ignore a large family of program constructs: features with pre- and postconditions whose only purpose is to serve as proof obligations. Such features do not implement any ADT functions and are not to be invoked. Instead, they are intended solely for static verification.

\begin{figure}[!b]
\begin{lstlisting}
extend_updates_item (s: STACK_IMPLEMENTATION [G]; x: G)
    do
        s.extend(x)
    ensure
        s.item = x
    end
\end{lstlisting}
\caption{Axiom A1 as a specified feature}
\label{fig:stack_a1_alternative}
\end{figure}

\autoref{fig:stack_a1_alternative} gives an example. The feature \e{extend_updates_item} is an alternative translation of axiom A1. It possesses the following properties:

\begin{itemize}
\item It operates on explicit objects \e{s} and \e{x}.
\item It uses an explicit invocation of implementation feature \e{extend}.
\end{itemize}

The example in \autoref{fig:stack_a1_alternative} takes the whole feature \e{extend_updates_item} as the translation of the axiom, as opposed to the one in \autoref{fig:stacks_traditional_approach}, where the axiom is captured with the assertion \e{item = x} in the postcondition of implementation feature \e{extend}.

\begin{figure}[!t]
\begin{lstlisting}
remove_then_extend (s1, s2: STACK_IMPLEMENTATION [G]; x: G)
    require
        s1.is_equal(s2)
    do
        s1.extend (x)
        s1.remove
    ensure
        s1.is_equal(s2)
    end
\end{lstlisting}
\caption{Axiom A2 as a specified feature}
\label{fig:stack_a2_alternative}
\end{figure}

Using this approach, it is possible to capture axiom A2 in the form of the feature \e{remove_then_extend} in \autoref{fig:stack_a2_alternative}. Again, the whole feature is the translation of the axiom. The feature \e{is_equal} defines an equivalence relation over run time objects representing stacks. It is declared by default in all Eiffel classes and compares its operands by value. The notion of equality deserves a separate analysis; \autoref{sec:details:implicit_proof_obligations} gives the details.

 Henceforth, this article will use the term \textbf{specification drivers} for specified features serving as translations of certain ADT axioms. A specification driver may be proven correct only if the implementation features it invokes have strong enough postconditions. Consequently, specification drivers, as their name suggests, drive specifying stronger postconditions.

\section{Specification Drivers in Practice}
\label{sec:details}

The present section derives the complete set of specification drivers for the stacks ADT (\autoref{fig:stacks_adt}). This set includes not only specification drivers that directly represent the original axioms of stacks because some specification drivers stem from a fundamental difference between ADT specifications and object-oriented programs: in the former it is not possible to have more than one occurrence of one and the same abstract stack, while in the latter it is possible to instantiate two run time objects denoting one and the same abstract stack. \autoref{sec:details:implicit_proof_obligations} and \autoref{sec:details:func_well_definedness} discuss the issue in detail and derive additional specification drivers caused by it.

\subsection{ADT axioms}
\label{sec:details:axioms}
Specification drivers do not bring any functional value to the system: they exist only to be eventually discharged as proof obligations. Consequently, they should not pollute implementation classes like \e{STACK_IMPLEMENTATION} in \autoref{fig:stacks_traditional_approach}. Concerning where to store them, the simplest option is to create a separate class within the source code project.
\begin{figure}[!b]
\begin{lstlisting}
deferred class ADT_AXIOMS_SPECIFICATION_DRIVERS [G]
feature {NONE}
	axiom_a1 (s: STACK_IMPLEMENTATION [G]; x: G)
		do
			s.extend (x)
		ensure
			s.item = x
		end

	axiom_a2 (s1, s2: STACK_IMPLEMENTATION [G]; x: G)
		require
			s1.is_equal (s2)
		do
			s1.extend (x)
			s1.remove
		ensure
			s1.is_equal (s2)
		end

	axiom_a3 (s: STACK_IMPLEMENTATION [G]; x: G)
		do
			s.extend (x)
		ensure
			not s.is_empty
		end

	axiom_a4: STACK_IMPLEMENTATION [G]
		do
			create Result.new
		ensure
			Result.is_empty
		end
end
\end{lstlisting}
\caption{Specification drivers capturing the axioms of stacks}
\label{fig:explicit_axioms}
\end{figure}
The \e{ADT_AXIOMS_SPECIFICATION_DRIVERS} class in \autoref{fig:explicit_axioms} contains specification drivers capturing the ADT axioms of stacks. This class is generic: since it talks about instances of a generic concept, \e{STACK_IMPLEMENTATION [G]} in this case, it needs to assume existence of type \e{G} to keep the genericity. The \e{\{NONE\}} clause suggests that the features listed within the corresponding \textbf{feature} block do not supply any useful functionality. The \e{deferred} keyword in front of the class declaration suggests that it is not possible to instantiate any objects of this class, which makes sense as the class serves as a document containing specification drivers rather than a blueprint for creating run time objects.
\subsection{Equivalence}
\label{sec:details:implicit_proof_obligations}
It is possible to see that the specification drivers in \autoref{fig:explicit_axioms} use two different operators for objects comparison: \e{=} and \e{is_equal}, while the original ADT specification in \autoref{fig:stacks_adt} invokes only $=$. This section discusses the difference between comparing instances of ADTs and comparing objects instantiated from object-oriented classes and introduces a set of specification drivers capturing the difference.

\begin{figure}[!t]
\begin{lstlisting}
s1, s2: STACK_IMPLEMENTATION [INTEGER]
create s1.new
create s2.new
\end{lstlisting}
\caption{Creating two instances of the empty stack}
\label{fig:two_empty_stacks}
\end{figure}

ADT specifications operate on sets of instances in the mathematical sense of the word "set": an abstract data type cannot contain two instances of one and the same abstract object. For example, the range of the function $new$ consists of the only stack instance, which is the empty stack, as axiom A4 suggests. When an object-oriented program is running, it is perfectly fine for it to have two run time objects in its memory denoting one and the same instance of the ADT. For example, it is possible to declare two variables of type \e{STACK_IMPLEMENTATION [INTEGER]} and make them both refer to two different stack objects in the memory, as in \autoref{fig:two_empty_stacks}. Consequently, run time objects form not a set of abstract objects, but a \textit{multiset}, or \textit{bag}~\cite{blizard1989multiset}. That is why there are two different comparison operators: the \e{=} operator checks whether the operands refer to identical run time objects, and \e{is_equal} checks whether the objects referenced by the operands represent the same instance of the ADT implemented by the class. As a consequence, if specification drivers representing ADT axioms use the feature \e{is_equal}, the corresponding implementation class should redefine the feature and its postcondition should be strong enough to satisfy the definition of equivalence relations.
\begin{figure}[!t]
\begin{lstlisting}
deferred class EQUIVALENCE_SPECIFICATION_DRIVERS [G]
feature {NONE}
	reflexivity (s: STACK_IMPLEMENTATION [G])
		do
		ensure
			s.is_equal (s)
		end

	symmetry (s1, s2: STACK_IMPLEMENTATION [G])
		require
			s1.is_equal (s2)
		do
		ensure
			s2.is_equal (s1)
		end

	transitivity (s1, s2, s3: STACK_IMPLEMENTATION [G])
		require
			s1.is_equal (s2)
			s2.is_equal (s3)
		do
		ensure
			s1.is_equal (s3)
		end
end
\end{lstlisting}
\caption{Capturing the definition of equivalence}
\label{fig:equivalence_definition}
\end{figure}
A relation over stacks is an equivalence relation if and only if it possesses the following properties:
\begin{itemize}
\item Reflexivity: every stack is equal to itself.
\item Symmetry: if stack $s_1$ is equal to stack $s_2$, then $s_2$ is equal to $s_1$ as well.
\item Transitivity: if stack $s_1$ is equal to stack $s_2$, and $s_2$ is equal to $s_3$, then $s_1$ is equal to $s_3$.
\end{itemize}
As \autoref{fig:equivalence_definition} illustrates, the three properties may be captured by a separate class created specifically for this goal. If all the features of class \e{EQUIVALENCE_SPECIFICATION_DRIVERS} are correct, then the postcondition of \e{is_equal} indeed defines an equivalence relation over run time objects instantiated from \e{STACK_IMPLEMENTATION [G]}.

It is worth noting that because equivalence definition is static, specification drivers for equivalence may be generated automatically for every class.

\subsection{Well-definedness}
\label{sec:details:func_well_definedness}
The ADT specification in \autoref{fig:stacks_adt} lists certain functions over stacks. It is necessary to ensure that they remain functions in the presence of an equivalence relation. Invoking a given implementation feature for two run time objects, which represent a single ADT object, should be indistinguishable from applying the ADT function implemented by this feature to that ADT object. Since a function application produces only one element from its range set, the two run time objects should also be considered equal after the invocation.
\begin{figure}[!t]
\begin{lstlisting}
deferred class WELL_DEFINEDNESS_SPECIFICATION_DRIVERS [G]
feature {NONE}
	new_is_well_defined (s1, s2: STACK_IMPLEMENTATION [G])
		require
			s1.is_empty
			s2.is_empty
			s1 /= s2
		do
		ensure
			s1.is_equal (s2)
		end

	is_empty_is_well_defined (s1, s2: STACK_IMPLEMENTATION [G])
		require
			s1.is_equal (s2)
			s1 /= s2
		do
		ensure
			s1.is_empty = s2.is_empty
		end

	item_is_well_defined (s1, s2: STACK_IMPLEMENTATION [G])
		require
			not s1.is_empty
			not s2.is_empty			
			s1.is_equal (s2)
			s1 /= s2
		do
		ensure
			s1.item = s2.item
		end

	extend_is_well_defined (s1, s2: STACK_IMPLEMENTATION [G]; x: G)
		require
			s1.is_equal (s2)
			s1 /= s2
		do
			s1.extend (x)
			s2.extend (x)
		ensure
			s1.is_equal (s2)
		end

	remove_is_well_defined (s1, s2: STACK_IMPLEMENTATION [G])
		require
			not s1.is_empty
			not s2.is_empty
			s1.is_equal (s2)
			s1 /= s2
		do
			s1.remove
			s2.remove
		ensure
			s1.is_equal (s2)
		end
end
\end{lstlisting}
\caption{Specification drivers for well-definedness}
\label{fig:well_definedness}
\end{figure}
This property is called \textbf{well-definedness} under an equivalence relation~\cite{dummit1991abstract}. The class \e{WELL_DEFINEDNESS_SPECIFICATION_DRIVERS} in \autoref{fig:well_definedness} contains specification drivers that encode well-definedness for every stacks implementation feature.  The \e{s1 /= s2} assertion in the preconditions emphasizes the fact that identical objects are of no interest in this context. Indeed, identity always implies equality, so in this case the well-definedness requirement is in fact tautological. The specification drivers \e{item_is_well_defined} and \e{remove_is_well_defined} also contain assertions \e{not s1.is_empty} and \e{not s2.is_empty}. These specification drivers invoke implementation features \e{item} and \e{remove} which have preconditions that need to be satisfied. The purpose of the mentioned assertions is exactly this.

Specification driver \e{new_is_well_defined} deserves special attention. In fact, it encodes something stronger than just the well-definedness of the implementation feature \e{new}. It says that two empty stacks are always equal. This makes perfect sense and at the same time implies the necessary well-definedness property: from the ADT specification in \autoref{fig:stacks_adt} and its first approximation in \autoref{fig:stacks_traditional_approach}, it is known that instantiating a stack with function $new$ results in the empty abstract stack. Consequently, the \e{new_is_well_defined} specification driver covers this case, since it applies to every pair of run time objects denoting the empty abstract stack.

Similarly to equivalence, the notion of well-definedness is long-established; as such, it is possible to generate the corresponding specification drivers automatically.

\subsection{Complete contracts}
\label{sec:details:complete_contract_example}

Although some works (\cite{polikarpova2014specified}, \cite{schoeller2006making}) talk about contract (in)completeness, they do not define this notion precisely. In light of the fundamental difference between ADT specifications and object-oriented programs, which causes the notion of equivalence over run time objects to appear (\autoref{sec:details:implicit_proof_obligations}), the definition cannot be implicitly equal to the definition of sufficiently complete ADT specifications~\cite{guttag1978algebraic} and needs to be written down explicitly.

As the other details of the original definition in~\cite{meyer1988object} do not bring any value to the discussion, this article uses a simplified definition of a contract.

\begin{definition}
A \textbf{contract} is a set composed of all pairs of the form $(Precondition(f), Postcondtion(f))$ for every implementation feature $f$.
\end{definition}

This definition ignores the possible presence of class invariants as it is always possible to get rid of them by appending to pre- and postconditions of the implementation features.

\begin{definition}
A contract is \textbf{correct} if and only if:
\begin{itemize}
\item Its postconditions are strong enough to ensure correctness of the specification drivers derived from the input ADT axioms (\autoref{sec:details:axioms})
\item In the event that specification drivers for the input ADT axioms use equivalence, its postconditions are strong enough to ensure correctness of the specification drivers for equivalence (\autoref{sec:details:implicit_proof_obligations}).
\end{itemize}
\end{definition}

\begin{definition}
A contract is \textbf{well-defined} if and only if its postconditions are strong enough to ensure correctness of the specification drivers for well-definedness (\autoref{sec:details:func_well_definedness}).
\end{definition}

\begin{definition}
A contract is \textbf{complete} if and only if it is correct and well-defined.
\end{definition}

\section{Related Work}
\label{sec:related_work}
Doctoral thesis~\cite{polikarpova2014specified} uses features with pre- and postconditions for checking completeness of model-based contracts (discussed later in this section). The definition of a complete model-based contract is not related to the definition of completeness in \autoref{sec:details:complete_contract_example}. According to~\cite{polikarpova2014specified}, completeness is what the current article calls well-definedness, expressed in terms of abstract mathematical concepts.

Although the specification driver approach allows capturing ADT axioms in their original form, it does specify how to actually build complete contracts having a set of specification drivers. As \autoref{sec:motivating_example} suggests, in many cases it is not possible to specify strong enough postconditions in terms of the ADT specification itself. This is where the need for representation appears: the implementation class has to stick to some already implemented data structure in order to enable stronger postconditions expressed it terms of this data structure. The problem of choosing an ideal representation has been aptly handled in multiple publications, therefore the present article does not propose its own methodology, but chooses instead to reference these publications.

Work~\cite{meyer2003framework} shows that it makes sense to use mathematical abstractions for representations: for example, it seems reasonable to think about stacks as mathematical sequences. That work also shows how to prove correctness against contracts strengthened with precise mathematical abstractions. Work~\cite{schoeller2006making} introduces the Mathematical Model Library (MML) - Eiffel library containing core abstractions: sets, sequences, bags, tuples etc. A more recent work~\cite{polikarpova2015fully} introduces EiffelBase2, a usable library of essential data structures, including stacks, represented as mathematical abstractions from MML. EiffelBase2 is fully verified with the AutoProof verifier. The underlying verification methodology~\cite{polikarpova2014flexible} assumes writing quite a number of assertions related to program execution semantics, so giving complete examples here would introduce confusion rather than clarity.
\begin{figure}[!t]
\begin{lstlisting}
class STACK_SEQUENCE_IMPLEMENTATION [G]
inherit	ANY redefine is_equal end
create	new -- Marking new as a creation feature
feature
	sequence: MML_SEQUENCE [G] -- Stack representation

	extend (x: G)  -- Extending with a new element
		do
		ensure
			a1: item = x
			a4: not is_empty
			definition: sequence = old sequence.extended (x)
		end

	remove -- Removing the topmost element
		require
			not is_empty
		do
		ensure
			definition: sequence = old sequence.but_last
		end

	item: G    -- Retrieving the topmost element
		require
			not is_empty
		do
		ensure
			definition: Result = sequence.last
		end

	is_empty: BOOLEAN  -- Is the stack empty?
		do
		ensure
			definition: Result = sequence.is_empty
		end

	new    -- Instantiating a stack
		do
		ensure
			a3: is_empty
			definition: sequence.is_empty
		end

	is_equal (other: STACK_SEQUENCE_IMPLEMENTATION [G]): BOOLEAN    -- Redefining equality
		do
		ensure then
			definition: Result = (sequence.count = other.sequence.count and then
						(across 1 |..| sequence.count as i all sequence[i.item] = other.sequence[i.item] end))

		end
end
\end{lstlisting}
\caption{Abstract model of stacks as sequences}
\label{fig:stacks_as_sequences}
\end{figure}
Instead, \autoref{fig:stacks_as_sequences} presents the idea in a nutshell. The \e{STACK_SEQUENCE_IMPLEMENTATION} class is the abstract model of stacks from the EiffelBase2 standpoint. EiffelBase2 equips classes implementing stacks with the \e{sequence} attribute and strengthens postconditions of the implementation features in terms of it. Class \e{MML_SEQUENCE} cannot be instantiated into any run time objects and exists only for verification purposes: it maps directly to the data structure representing mathematical sequences in the underlying proving engine. The \e{sequence} attribute is further connected to meaningful data structures by means of abstraction and refinement techniques~\cite{hoare2002proof}. Works~\cite{polikarpova2015fully} and~\cite{polikarpova2014specified} give more implementation details.

 In \autoref{fig:stacks_as_sequences}, the implementation features are formally defined with assertions over the \e{sequence} attribute (marked with the "definition" tag) added to the features' postconditions. The comparison feature \e{is_equal} is redefined so that two stacks are considered equal if and only if the sequences representing them are equal. Two sequences are considered equal if and only if their sizes are equal and they contain same objects. The feature \e{extended} models a sequence where an object is appended to the target sequence on to which the feature is invoked; feature \e{but_last} models the target sequence, but without the last element; feature \e{last} models the element added to the target sequence last; feature \e{is_empty} models the indication whether the target sequence is empty or not; finally, feature \e{count} models the size of the target sequence.

Mathematical concepts from MML are abstract, but they still form particular representations in EiffelBase2, though mathematically precise. The concept of model-based contracts helps to specify complete contracts, but does not say how to rigorously check contracts with representations for completeness. Furthermore, it fails to define what complete contracts are. The notion of specification drivers bridges this gap. All the specification drivers derived in the present article are expressed in terms of the original ADT specification (\autoref{sec:details:axioms}) plus the abstract equivalence (\autoref{sec:details:implicit_proof_obligations} and \autoref{sec:details:func_well_definedness}), whose presence is inevitable due to the nature of computing which allows programs to keep in their memory several instances of one and the same abstract object. They do not require making any assumptions about possible representations and enable defining complete contracts precisely.

\section{Proving contracts completeness}
\label{sec:proving_completeness}

It is possible to give a manual proof of completeness of the contract depicted in \autoref{fig:stacks_as_sequences}. Fortunately, this work may be done automatically. This advantage makes it possible to apply the specification drivers approach to legacy implementations. Indeed: if there is a source code project with a number of classes in it, then it is possible to devise an additional class, write all the applicable specification drivers into it and submit the resulting class to the prover. Instead of showing how to derive complete contracts having a set of specification drivers from scratch, the article shows how to apply the approach to existing contracts.

The EiffelBase2 library seems to be a natural choice for the experiment. The library contains a complete implementation of stacks specified as mathematical sequences. The corresponding implementation class is \e{V_LINKED_STACK}. In order to perform the experiment, it is necessary to take the stacks specification drivers from \autoref{sec:details} and modify them so that the name of the implementation class would be \e{V_LINKED_STACK} instead of \e{STACKS_IMPLEMENTATION}. The specification driver \e{axiom_a4} comes with a pitfall: the \e{V_LINKED_STACK} class does not introduce its own creation feature, but redefines the default creation feature defined for all classes. Hence, the \e{create Result.new} instruction is not applicable here; one should use \e{create Result} instead. After these modifications, the specification drivers should successfully compile and be ready for verification.

The initial verification attempt using AutoProof will result in numerous precondition violations. As \autoref{sec:related_work} suggests, the verification methodology~\cite{polikarpova2014flexible} behind AutoProof assumes writing additional non-stack related assertions. For example, the \e{extend_is_well_defined} specification driver can be verified by AutoProof only in the form depicted in \autoref{fig:driver_for_autoproof}.
\begin{figure}[!t]
\begin{lstlisting}
extend_is_well_defined (s1, s2: V_LINKED_STACK [G]; x: G)
	require
		s1.is_wrapped
		s2.is_wrapped
		s1.observers.is_empty
		s2.observers.is_empty
		modify([s1, s2])


		s1.is_equal (s2)
		s1 /= s2
	do
		s1.extend (x)
		s2.extend (x)
	ensure
		s1.is_equal (s2)
	end
\end{lstlisting}
\caption{Specification driver for verifying by AutoProof}
\label{fig:driver_for_autoproof}
\end{figure}
The five assertions in the beginning of the \e{require} precondition clause seem to be worth explaining them briefly. The \e{s1.is_wrapped} assertion says that reference \e{s1} is assumed to be non-void and not participating in any call; the \e{s1.observers.is_empty} assertion says that the set of objects interested in the state of \e{s1} should be empty - it is a part of the precondition of feature \e{extend} of class \e{V_LINKED_STACK}; finally, the \e{modify([s1, s2])} assertion is a frame specification: it says that the enclosing feature, \e{extend_is_well_defined} in this case, is going to modify objects referenced by \e{s1} and \e{s2} (square brackets \e{[]} denote set constants in Eiffel). The precondition needs the \e{modify} assertion because the \e{extend_is_well_defined} feature uses feature invocations with side effects, \e{extend} in this case, on references \e{s1} and \e{s2}. Although the verification failures caused by the absence of these assertions do not bear any relation to stacks, they uncover certain weaknesses in the verification methodology: namely, the defaults do not seem sufficiently reasonable. For example, a violation of the \e{s1.is_wrapped} assertion would detect a callback situation, and callbacks are not so common as to assume them by default. The \e{observers.is_empty} requirement makes extending stack objects applicable only in situations when no other objects depend on their states. The \e{modify} frame specification may be generated automatically based on the presence of invocations with side effects in the implementation body.

After complementing the specification drivers with all necessary assertions related to verification methodology and rerunning AutoProof, it uncovers some stack-related issues. This is visible from the fact that this time the verification errors come from the postconditions. Namely, AutoProof fails to prove correctness of all the verification drivers from classes \e{EQUIVALENCE_SPECIFICATION_DRIVERS} and \e{WELL_DEFINEDNESS_SPECIFICATION_DRIVERS} as well as verification driver \e{axiom_a2} from the \e{ADT_AXIOMS_SPECIFICATION_DRIVERS} class. As all of these specification drivers involve implementation feature \e{is_equal}, the first guess is that \e{V_LINKED_STACK} does not redefine it. This guess appears to be right: the class defines its own custom feature for comparing run time objects, but does not redefine the standard comparison feature in terms of the new one. Giving this flaw's fix here would not bring much value to the discussion, so it seems better to move on. After redefining feature \e{is_equal}, AutoProof succeeds in proving classes \e{ADT_AXIOMS_SPECIFICATION_DRIVERS} and \e{EQUIVALENCE_SPECIFICATION_DRIVERS} completely, but still fails to prove specification driver \e{new_is_well_defined} from the \e{WELL_DEFINEDNESS_SPECIFICATION_DRIVERS} class. As this specification driver uses the \e{is_empty} implementation feature, it falls under suspicion. Apparently, its postcondition does not have a clause corresponding to the \e{definition} clause in its abstract model in \autoref{fig:stacks_as_sequences}. After fixing this flaw, everything verifies successfully, including the \e{V_LINKED_STACK} implementation class.

\section{Conclusions and further work}
\label{sec:conclusions}
The article makes the following main contributions:

\begin{itemize}
\item Presents the specification driver approach for encoding ADT axioms, which are not possible to encode using traditional DbC techniques.
\item Illustrates the process of axiomatizing abstract equivalence using the new approach.
\item Introduces an exhaustive definition of contract completeness.
\item Demostrates how to apply completeness checks to legacy implementations.
\end{itemize}

The new approach allows adding, changing or removing ADT axioms at any given moment of the development process without necessarily modifying the implementation classes. Although specification drivers occupy separate classes completely disjoint from implementation classes, they are simultaneously expressed in terms of objects instantiated from the implementation classes. The result is a seamless integration of software axiomatization and implementation driven by automatic verification of functional correctness. Attempts to check specification drivers can uncover weak postconditions of implementation features. Once strengthened, these postconditions potentially yield firmer executable instructions.

In light of the presence of different kinds of specification drivers described in \autoref{sec:details} it seems feasible to propose the following changes to the Eiffel Verification Environment tool set:

\begin{itemize}
\item Develop a template for fast creation of classes intended to keep specification drivers.
\item Automate generation of specification drivers for equivalence and well-definedness.
\item Revise verification methodology underlying AutoProof: in essence, specification drivers are a new syntactical specification construct, which may potentially remove some particularly egregious verification challenges.
\end{itemize}
Work~\cite{Meyer13Multi} introduces the notion of multirequirements, and work~\cite{anaumchev2015seamless} illustrates how to apply this notion in practice. The underlying idea is that a separate item in a software requirements document should be expressed using several interwoven notations, e.g. natural language, graphical form and formal notation. For the formal notation, it was suggested to use a rather expressive programming language. The present paper talks about expressing ADT axioms in a programming language with pre- and postconditions. Since ADT specifications are one of the languages for expressing software requirements, it makes sense to revisit the original multirequirements approach to see how the idea of specification drivers could improve it.

The idea of specification drivers was inspired mostly by seminal works~\cite{hoare2002proof} and~\cite{meyer1988object}, and driven by the will to unify requirements and code seeded in the work~\cite{Meyer13Multi}.


\section*{Acknowledgment}
The authors would like to thank Innopolis University for supporting the Software Engineering Laboratory where the research resulted in this work is taking place.

Special thanks goes to Daniel Johnston from MSIT-SE Program at Innopolis University, who kindly agreed to proofread this paper line-by-line.



%
{{{
\bibliographystyle {IEEEtran}
\bibliography {SpecificationDrivers_v2}
}}}

\end{document}